\documentclass[a4paper]{article}
\usepackage[margin=1in]{geometry}
\usepackage{booktabs}
\usepackage{authblk}
\usepackage{amsmath}
\usepackage{csvsimple}

\usepackage{caption}
\usepackage{subcaption}

\usepackage{rotating}
\usepackage{tikz}

\usepackage{graphicx}

\usepackage{anyfontsize}

\usepackage{titlesec}
\titleformat{\chapter}[display]{\huge\bfseries}{\chaptertitlename~\thechapter}{20pt}{\Huge}
\titlespacing{\chapter}{0pt}{50pt}{40pt}
\title{{Energy Deposition due to Secondary Particles in The Helical Undulator Vacuum at ILC-250GeV*}}

\author[1,2,3]{K. Alharbi} 
\author[3]{S. Riemann} 
\author[1]{A.Alrashdi}
\author[2,4]{G. Moortgat-Pick} 
\author[2]{A. Ushakov }

\affil[1]{\textit{King Abdul-Aziz City for Science and Technology (KACST), Riyadh, Kingdom of Saudi Arabia}}
\affil[2]{\textit{University of Hamburg, Luruper Chaussee 149, D-22761 Hamburg, Germany}}
\affil[3]{\textit{Deutsches Elektronen-Synchrotron (DESY), Platanenallee 6, D-15738 Zeuthen, Germany}}
\affil[4]{\textit{Deutsches Elektronen-Synchrotron (DESY), Notkestrasse 85, D-22607 Hamburg, Germany}}

\date{}
\begin{document}\maketitle

\begin{abstract}

\fontsize{13}{13}\selectfont

In the future, International Linear Collider (ILC), a helical undulator based polarized positron source is expected to be chosen. A high energy electron beam passes through a superconducting helical undulator in order to create circularly polarized photons which will be directed to a conversion target, the result of which, will be electron-positron pairs. The resulting positron beam is longitudinally polarized. In order to produce the required number of positrons in ILC250 the full undulator length is needed. Since the photons are created with an opening angle and traveling through a 320 m long undulator, it is expected that the superconducting undulator vacuum will be hit by the photons. Photon masks are needed to be inserted in the undulator line to keep the power deposition in the vacuum below the acceptable limit which is 1W/m. A detailed study of the power deposition in the vacuum and masks is needed in order to design the photon masks. This paper describes the power deposition in the undulator vacuum due to secondary particles assuming an ideal undulator. In addition, the mask model is proposed.

\end{abstract}

\begin{flushleft}
\rule{200pt}{0.5pt}\\
\fontsize{10}{10}\selectfont
{"Talk presented at the International Workshop on Future Linear Colliders (LCWS2019), Sendai, Japan, 28 October-1 November, 2019. C19-10-28."\\
khaled.alharbi@desy.de}
\end{flushleft}

\newpage

\fontsize{12}{22}\selectfont
\section{Introduction}
A helical undulator in the International linear collider (ILC) is anticipated to be chosen as the baseline positron source. The helical undulator produces polarized positrons with the greatest intensity that can be achieved by any available intense sources for the time being. An electron beam with 128 GeV that comes from the main linac will enter the helical undulator in order to produce circularly polarized photons. These multi-MeV photons will then hit a thin Ti-alloy target. As a result, longitudinally polarized positrons will be produced through pair production mechanism. These positrons will be collected, accelerated and sent to the damping ring. 
Table  \ref{table1} shows the ILC undulator parameters. Technical Design Report [TDR] describes the general layout of the helical undulator \cite {adolphsen2013international}. Two helical undulator modules with a length of 1.75 m for each are installed in one cryomodule which has a length of 4.1 m.  23 quadrupoles are placed inside the helical undulator to steer and focus the electron beam through the helical undulator. The distance between each two quadrupoles is 14.538 m. The saved area for the undulator in ILC-250GeV is 320 m whearas the active length of the helical undulator is 231m. The undulator aperture and period are 5.85 and 11.5 mm, respectively. The maximum B field on helical undulator axis is 0.86T corresponding to a maximum K value of 0.92. Here, K=0.85 was used.

\begin{table}[h]
\caption[ILC undulator parameters]{ILC undulator parameters.}
\centering 
\begin{tabular}{l*{6}{l}r} 
\hline\hline 
Parameters & Values \\ [0.5ex] 
\hline 
Centre-of-mass energy  & 250 GeV \\ 
Undulator period & 11.5 mm\\
Undulator K & 0.85 \\
Electron number per bunch & $2\times10^{10}$ \\
Number of bunches per pulse & 1312 \\
Pulse rate &  5.0 Hz \\
Cryomodule Length & 4.1 m \\
Effective magnet length & 3.5 m \\
Undulator aperture & 5.85 mm \\
Number of quadrupoles  & 23 \\ 
Quadrupole spacing & 14.538 m\\
Quadrupole length & 1 m \\
Total active undulator length & 231 m \\
Total lattice length & 319.828 m \\

\hline 
\end{tabular}
\label{table1} 
\end{table}

\section{Power Deposited at Undulator Vacuum}
Photons produced by the helical undulator will hit the helical undulator vacuum chamber; besides power deposition in the vacuum, the vacuum quality can also be affected~\cite {malyshev2007vacuum}. 
The power deposition due to synchrotron radiation along the helical undulator at ILC-250GeV was studied \cite {alharbi2019energy}. It showed that the peak power deposited due to the synchrotron radiation is about 18.5 W/m \cite {alharbi2019energy} while the acceptable limit of the power deposition in the undulator vacuum is 1W/m \cite {scott2008investigation}. Therefore photon masks should be placed into the undulator line to protect the undulator vacuum.

\section{The Effects of Inserting Photon Masks into the Undulator} 
Since we will never have ideal photon masks, secondary particles that have enough energy to escape from masks may deposit on the undulator vacuum. In addition, since the polarization of positron depends on the polarization of photons, the masks could affect the polarization of the positron beam by removing the low energy photons which have large opening angles. 

In this paper the power deposition due to secondary particles in the undulator vacuum will be discussed. Moreover, a possible photon mask model is proposed

\subsection{Energy Deposited at Photon Masks}

The energy deposited at masks was calculated \cite {alharbi2019energy} by an analytical equation and simulated by HUSR \cite {newton2010rapid} and \cite {newton2010modeling}. Table \ref{table2} summarizes this study. It shows the deposited power and the average and maximum incident photon energy at photon masks.

\begin{figure}[h]
\centering
\includegraphics[scale=0.6]{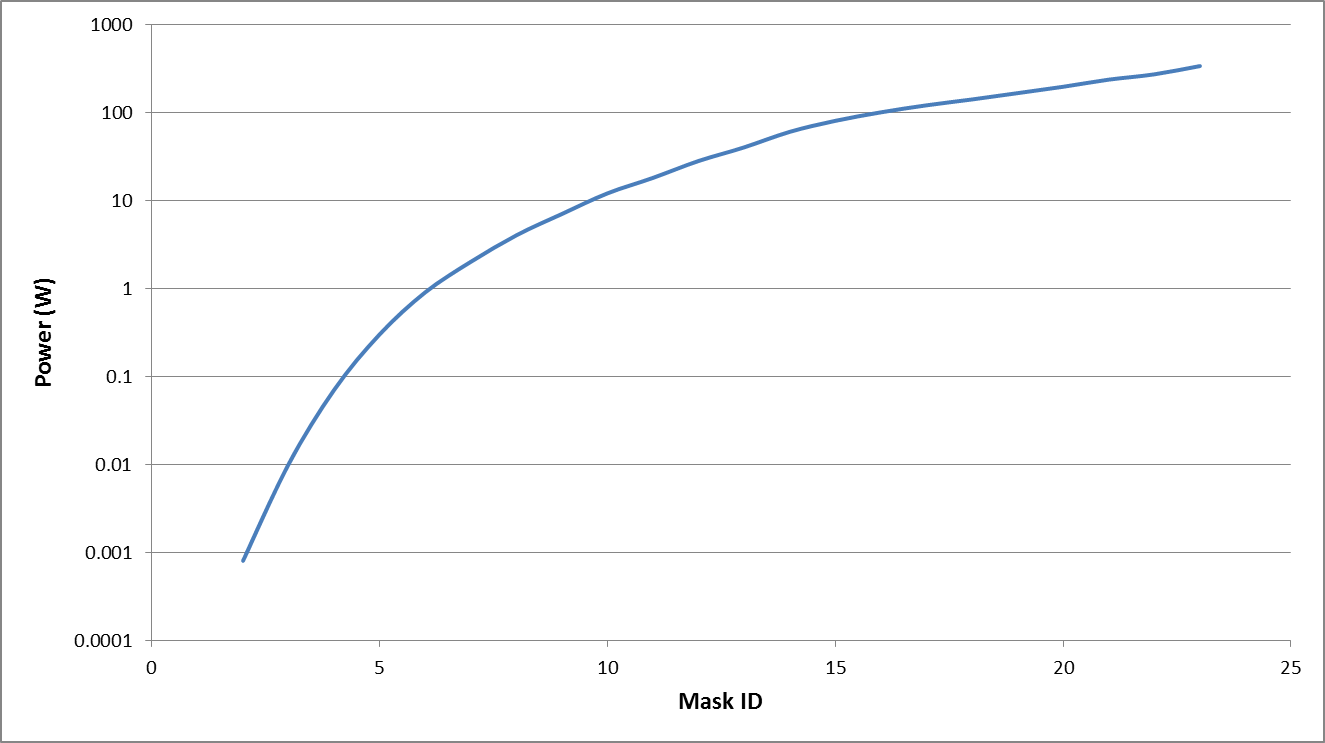}
\caption{Power Deposition at Photon Masks \cite {alharbi2019energy}.}
\label{fig:graph1}
\end{figure}

Figure \ref{fig:graph1} shows the power at the 22 masks. The first mask is missed due to the fact that HUSR is too time consuming to calculate the photon spectrum for distances below 25 m from the exit of an undulator module \cite {alharbi2019energy}. It can be expected that as shown in figure \ref{fig:graph1} , the first mask (Mask 1) will receive less than 0.001 W/m which the second mask (Mask 2) receives. So it is negligible.

\begin{table}[h]
\caption[ILC undulator parameters]{Average and maximum incident photon energy and power deposited at masks from Mask 15 to mask 23 in ILC-250 GeV \cite {alharbi2019energy}.}
\centering 
\begin{tabular}{l*{6}{l}r} 
\hline\hline 

Mask ID &       Average Incident Photon &        Maximum Incident Photon & Power (W)\\
&  Energy (MeV)& Energy (MeV) &\\
\hline 

15 &  0.68 &  3.49 & 80\\
16 & 0.85 & 4.39 & 100\\
17 & 0.97 &  5.11 & 120\\
18 & 1.13 & 6.16 & 140\\
19 &  1.31 &  6.98 & 165\\
20 &  1.46 & 7.85 & 195\\
21 &  1.65 & 8.97 & 235\\
22 &  1.77 & 9.81 & 270\\
23 & 2.01 & 10.77 & 335\\

\hline 
\end{tabular}
\label{table2} 
\end{table}

\section{Photon Mask Design}

The photon mask for the ILC helical undulator has not been completely designed  so far. And to design a mask, the distribution of the deposited energy is needed. It is clear that as shown in figure  \ref{fig:graph1} each mask will receive a different deposited power. For example, mask number 23 receives the highest amount of the power deposited which is 335 W compared with 80 W at mask number 15. Moreover, the average incident photon energy at mask 23 is about 2.01 MeV so it is above the threshold  which causes the pair production mechanism. While the average incident photon energy at mask 15 is 0.68 MeV. Therefore, photon masks should absorb the power deposited. 

Table  \ref{table2} shows the deposited power, average and maximum incident photon energy at the last 9 masks in the undulator (from Mask 15 to Mask 23). However the average incident photon energy is about 0.68 MeV at mask 15 and some photons carry energy up to 3.5 MeV. In addition the average incident photon energy at mask 23 is 2.01 MeV with some of the photons carrying energy up tp 10.77 MeV. Therefore the mask should be designed to absorb photons with different energies as well as the electromagnetic shower. 

Previously, a design of the photon mask was proposed by Bangau  \cite {bungau2008design} but  with different parameters concerning the electron energy and undulator K and total  length.\\

\subsection{Dimensions of the Photon Mask}

\begin{figure}[h]
\centering
\includegraphics[scale=1.2]{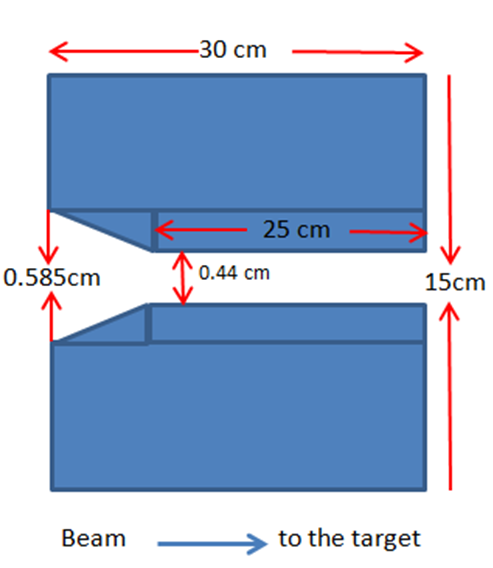}
\caption{Schematic layout of the mask.}
\label{fig:graph12}
\end{figure}

Figure  \ref{fig:graph12} shows the design proposed of the photon mask in this paper. Since the 
space reserved for the undulator in ILC250 GeV is limited (319.8m), the length of 
the mask has been chosen to be 30 cm. The outer radius of the mask is 15 cm. 
The inner radius of the photon mask is 0.44 cm. In order to reduce the effect of 
wakefields, the inner radius of the photon mask in the first 5 cm is tapered from 
0.585 cm to 0.44 cm.


\subsection{Material of the Photon Mask}

The incident photon energy at masks can reach few MeV. Therefore a material with small radiation length and high atomic number and high density is required.

Three different materials were investigated in this paper including copper, iron and tungsten. Table \ref{table13} summarizes the properties of these materials. The atomic number of tungsten is 74 whereas it is 26 and 29 for iron and copper, respectively. The density of tungsten is 19.3 g/$cm^3$ and is higher than the dinsity of copper and iron which are 8.96 and 7.89, respectively. For the melting point, it is 1357.77 K and 1811 K in case of copper and iron, respectively, while the melting point for tungsten is 3695 K. Another important property for these candadites is the radiation length. It is 1.436 cm and 1.757 cm for copper and iron, respectively, but tungsten has only 0.34 cm of radiation length.\\
Since the different materials are used, it is expected to see some differences in terms of the peak energy deposition density (PEDD), maximum temperature rise and the flux of secoundary particles that leave the photon mask.

\begin{table}[h]
\caption[ILC undulator parameters]{Properties of the chosen materials.}
\centering 
\begin{tabular}{l*{6}{l}r} 
\hline\hline 

Parameter &              Unit  & Copper & Iron & Tungsten\\

\hline 

Atomic Number && 29 & 26 & 74 \\
Density & g/$cm^3$  & 8.96 &   7.89    & 19.3    \\
Thermal Conductivity   &  W/(m.K)& 401 & 83.5 &          182           \\
Heat Capacity &  J/g/K  &  0.385 &   0.45   &  0.134   \\
Melting Point &  K  & 1357.77 &   1811 &   3695\\
Radiation Length & cm  & 1.436 &  1.757 & 0.35 \\

\hline 
\end{tabular}
\label{table13} 
\end{table}

\section{FLUKA Simulation}

FLUKA Monte Carlo code for particle tracking and particle interactions with matter \cite {fasso2005fluka} has been used to simulate the energy deposition in the photon mask with different materials.

\subsection{Energy Deposited at Copper Mask}

\begin{figure}[ht]
\centering

\begin{subfigure}{0.5\textwidth}
\centering
\includegraphics[scale=0.7]{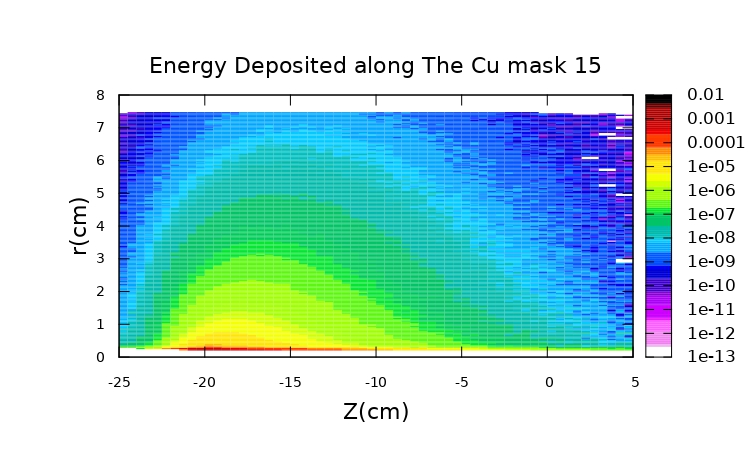}
\label{fig:graph8}
\end{subfigure}

\begin{subfigure}{0.5\textwidth}
\centering
\includegraphics[scale=0.7]{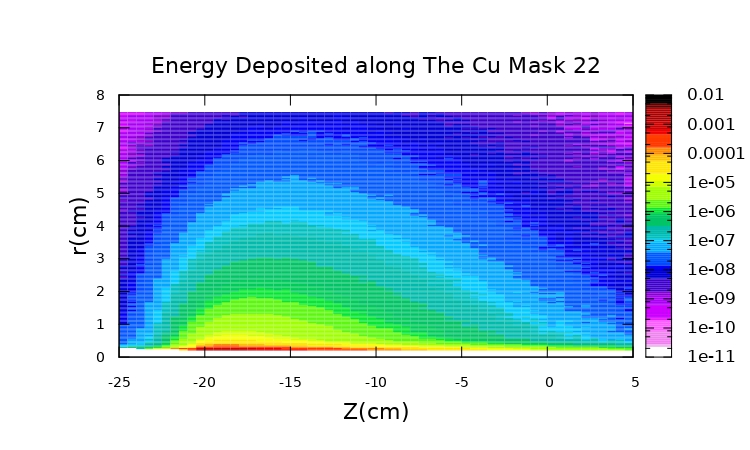}
\label{fig:graph9}
\end{subfigure}

\begin{subfigure}{0.5\textwidth}
\centering
\includegraphics[scale=0.7]{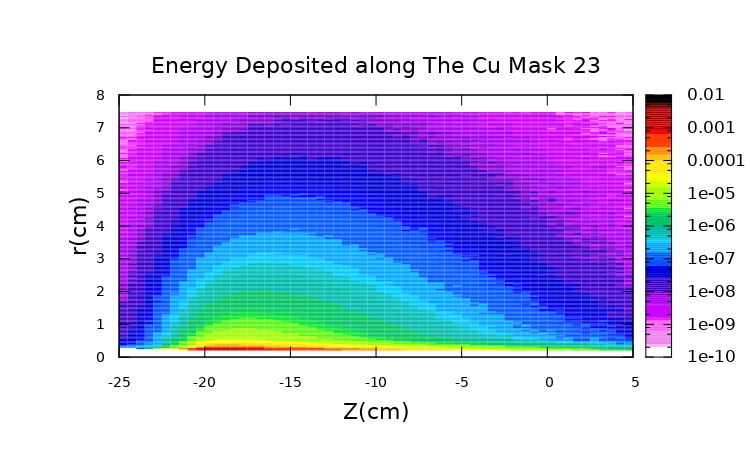}
\label{fig:graph9}
\end{subfigure}

\caption{Distribution of the energy deposited along the copper photon mask number 15, 22 and 23 from the top to the bottom, respectively, in 2D. }
\label{mul_images3}
\end{figure}

\begin{figure}[ht]
\centering

\begin{subfigure}{0.5\textwidth}
\centering
\includegraphics[scale=0.7]{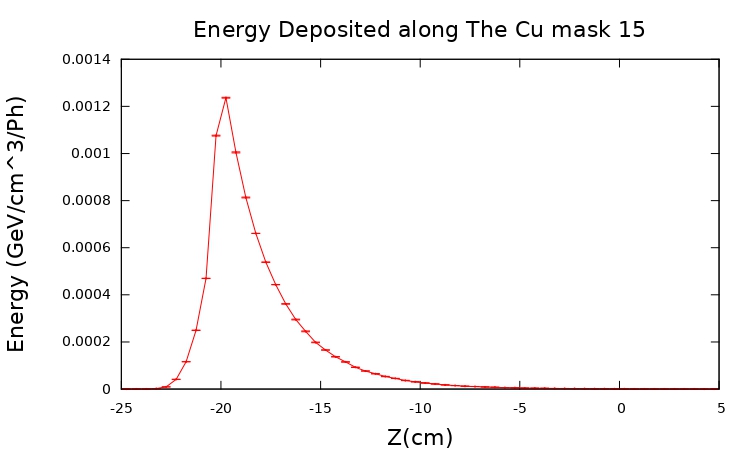}
\label{fig:graph8}
\end{subfigure}

\begin{subfigure}{0.5\textwidth}
\centering
\includegraphics[scale=0.7]{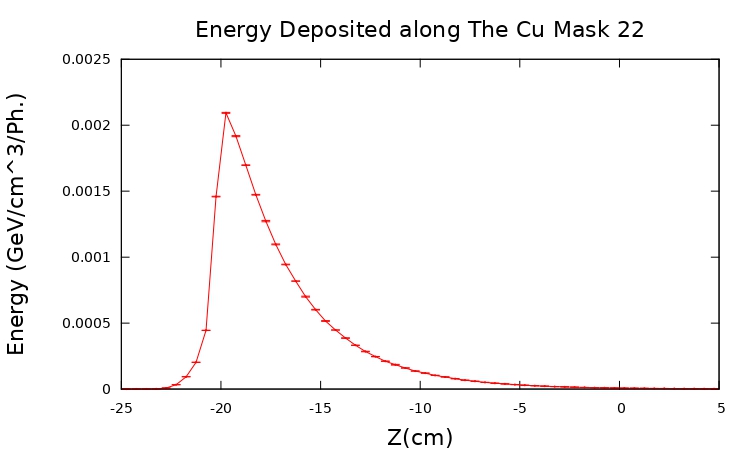}
\label{fig:graph9}
\end{subfigure}

\begin{subfigure}{0.5\textwidth}
\centering
\includegraphics[scale=0.7]{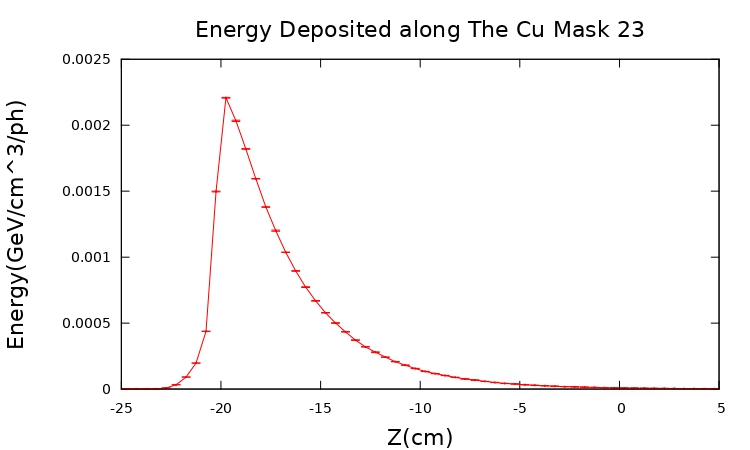}
\label{fig:graph9}
\end{subfigure}

\caption{Distribution of the deposited energy along the copper photon mask number 15, 22 and 23 from the top to the bottom, respectively, in 1D.}
\label{mul_images4}
\end{figure}

Figure \ref{fig:graph1} illustrates the amount of the power deposition in photon masks due to the synchrotron radiation. Figures \ref{mul_images3} and \ref{mul_images4} show the simulation of the energy deposited along three different copper masks, including mask number 15, 22 and 23 in 2D and 1D, respectively.\\

\newpage

\newpage

Table \ref{table14} shows the PEDD, maximum temperature rise and the power stopped by copper masks from mask 15 to mask 23. It is clear that the lowest PEDD is at mask number 15 while the highest PEDD is at mask number 23. The maximum temperature rise is 21.25 K/pulse at mask number 23, in contrast the lowest maximum 
temperature rise is 8.1 K/pulse at mask 15.\\ 
In general, copper masks can stop between 98.1$\%$ and 98.5 $\%$ of the incident power. For example, copper mask 23 can stop up to 98.5$\%$ of the incident power. The power that leaves the copper mask will be discussed in 5.4.

\begin{table}[h]
\caption[ILC undulator parameters]{PEDD, maximum temperature rise and power stopped by the last 9 copper masks (from mask 15 to mask 23).}
\centering 
\begin{tabular}{l*{6}{l}r} 
\hline\hline 

Mask ID &  Power   &    Average incident   & Energy  & PEDD& Maximum  & Stopped \\
 &   Deposited  &     photon energy  &  deposited & &  temperature &  Power\\
 &  [W]   &    [MeV]   & [GeV/cm$^3$/Ph]  & [J/(g*pulse)]& [K/pulse]  & $\%$ \\
\hline 

15 & 80   &  0.68 &   0.00123    &  3.12      &      8.1   &         98.1             \\
16 & 100 & 0.85  &   0.00138    &  3.63      &      9.43      &      98.1                 \\ 
17 & 120 &  0.97 &   0.00150    &  4.02      &    10.44      &           98.1            \\
18 & 140 &  1.13 &   0.00163    &  4.47      &      11.6         &        98.2             \\
19 & 165 &  1.31 &   0.00178    &  5           &       12.99       &        98.3                \\
20 &195  & 1.46  &   0.00186    & 5.55      &        14.4        &         98.4           \\
21 & 235 & 1.65  &   0.00198    & 6.29      &       16.35       &           98.5             \\
22 & 270 & 1.77  &   0.00210    & 7.15      &         18.57      &           98.5            \\
23 & 335 & 2.01  &   0.00220    & 8.18      &          21.25     &              98.5          \\

\hline 
\end{tabular}
\label{table14} 
\end{table}


\subsection{Energy Deposited at Iron Mask}
\begin{figure}[ht]
\centering

\begin{subfigure}{0.5\textwidth}
\centering
\includegraphics[scale=0.7]{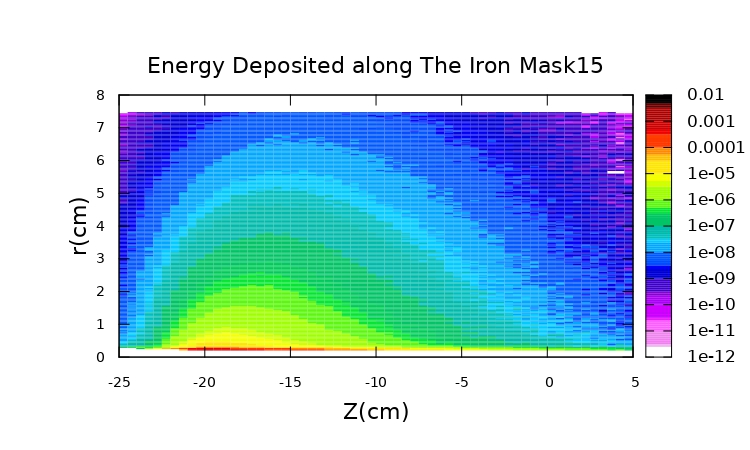}
\label{fig:graph8}
\end{subfigure}

\begin{subfigure}{0.5\textwidth}
\centering
\includegraphics[scale=0.7]{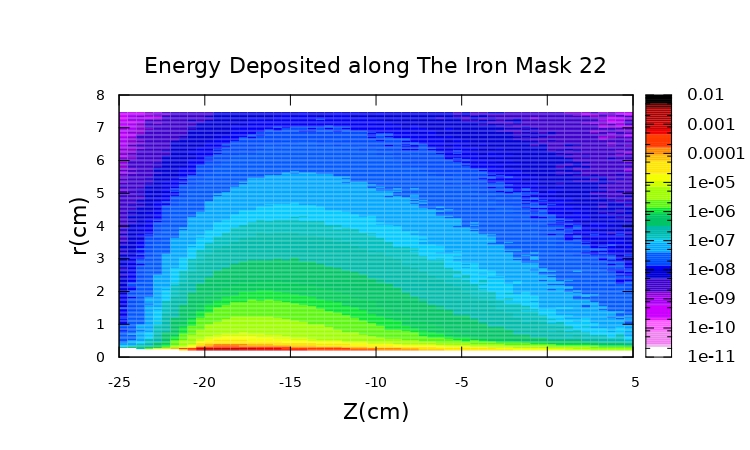}
\label{fig:graph9}
\end{subfigure}

\begin{subfigure}{0.5\textwidth}
\centering
\includegraphics[scale=0.7]{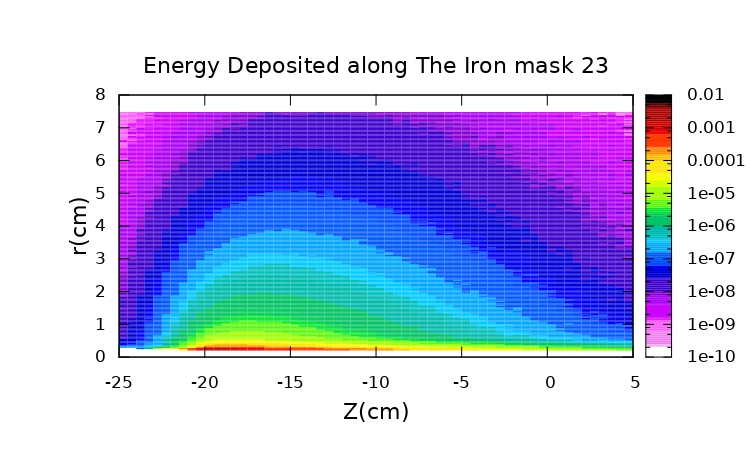}
\label{fig:graph9}
\end{subfigure}

\caption{Distribution of the energy deposited along the iron photon mask number 15, 22 and 23 from the top to the bottom, respectively, in 2D. }
\label{mul_images5}
\end{figure}

\begin{figure}[ht]
\centering

\begin{subfigure}{0.5\textwidth}
\centering
\includegraphics[scale=0.7]{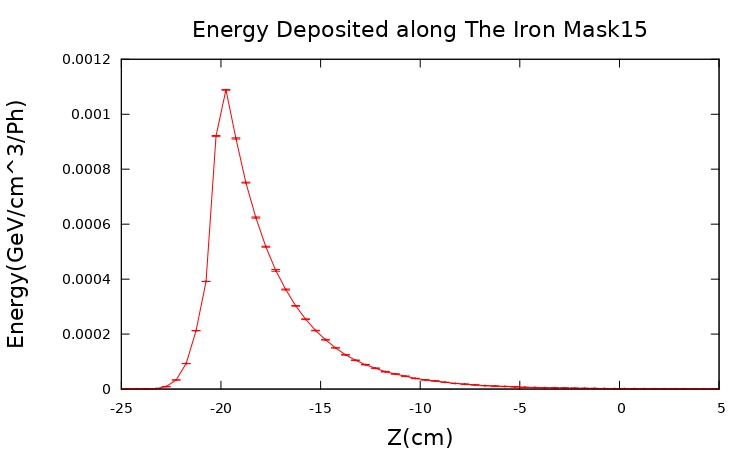}
\label{fig:graph8}
\end{subfigure}

\begin{subfigure}{0.5\textwidth}
\centering
\includegraphics[scale=0.7]{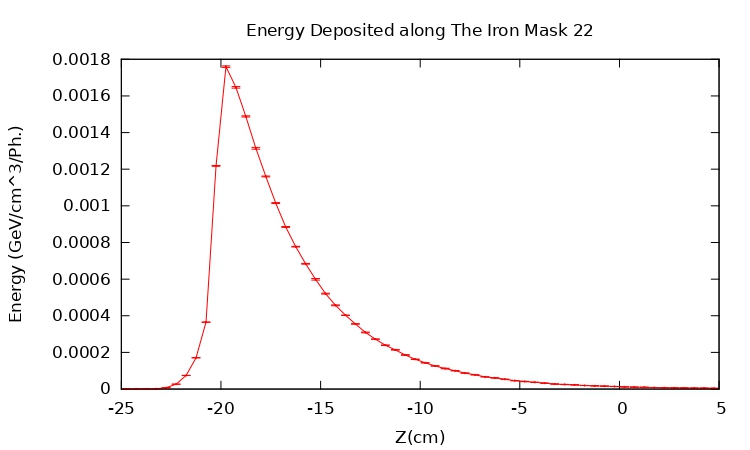}
\label{fig:graph9}
\end{subfigure}

\begin{subfigure}{0.5\textwidth}
\centering
\includegraphics[scale=0.7]{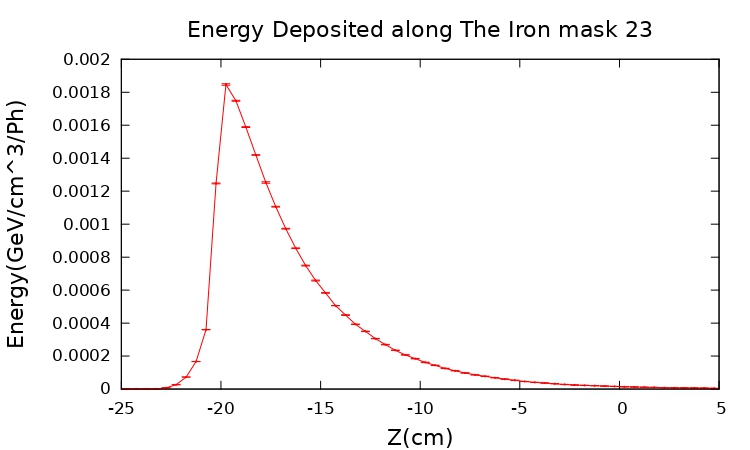}
\label{fig:graph9}
\end{subfigure}

\caption{Distribution of the energy deposited along the iron photon mask number 15, 22 and 23 from the top to the bottom, respectively, in 1D.}
\label{mul_images6}
\end{figure}

The second candidate is iron. Iron has a lower density than copper but a higher melting point. The energy deposited at iron masks number 15, 22 and 23 is shown in figure \ref{mul_images5} and \ref{mul_images6} in 2D and 1D, respectively.

Table \ref{table15} shows the PEDD, maximum temperature rise and the power stopped by iron masks. It is clear that the lowest PEDD is at mask number 15. While the highest PEDD is at mask number 23. It is 7.79 J/g/pulse while the PEDD of mask 23 is 7.79 J/g/pulse. The maximum temperature rise is 17.55 K/pulse at mask number 23, in contrast the lowest maximum temperature rise is 7.15 K/pulse at mask 15. Iron masks can stop between 97.3 $\%$ and 97.5 $\%$ of the incident power. For example, iron mask 23 can stop up to 97.5 $\%$ of the incident power. The power that leaves the iron mask will be discussed in 5.4.

\newpage

\newpage

\newpage
\begin{table}[h]
\caption[ILC undulator parameters]{PEDD, maximum temperature rise and power stopped by the iron mask number 15, 22 and 23.}
\centering 
\begin{tabular}{l*{6}{l}r} 
\hline\hline 

Mask ID &  Power   &    Average incident   & Energy  & PEDD& Maximum  & Stopped \\
 &   Deposited  &     photon energy  &  deposited & &  temperature &  Power\\
 &  [W]   &    [MeV]   & [GeV/cm$^3$/Ph]  & [J/(g*pulse)]& [K/pulse]  & $\%$ \\
\hline 
 
15 & 80   &  0.68 &   0.00123    &  3.18      &         7.15       &              97.3             \\
22 & 270 & 1.77  &   0.00210    & 6.78      &      15.28     &              97.5             \\
23 & 335 & 2.01  &   0.00220    &7.79      &         17.55    &              97.5               \\

\hline 
\end{tabular}
\label{table15} 
\end{table}


\subsection{Energy Deposited at Tungsten Mask}

\begin{figure}[ht]
\centering

\begin{subfigure}{0.5\textwidth}
\centering
\includegraphics[scale=0.7]{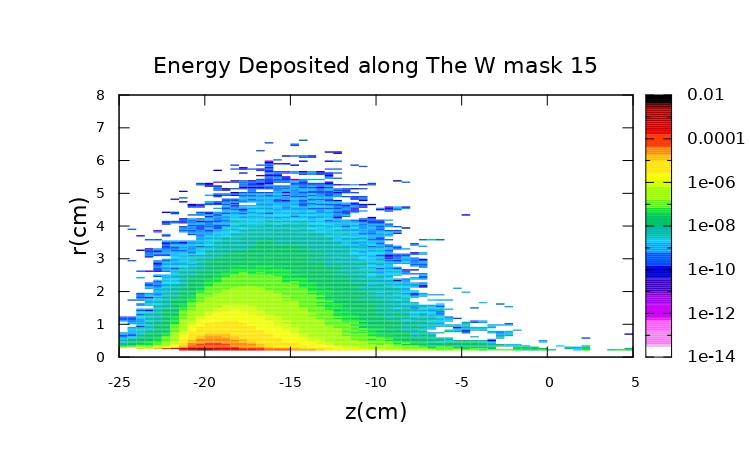}
\label{fig:graph8}
\end{subfigure}

\begin{subfigure}{0.5\textwidth}
\centering
\includegraphics[scale=0.7]{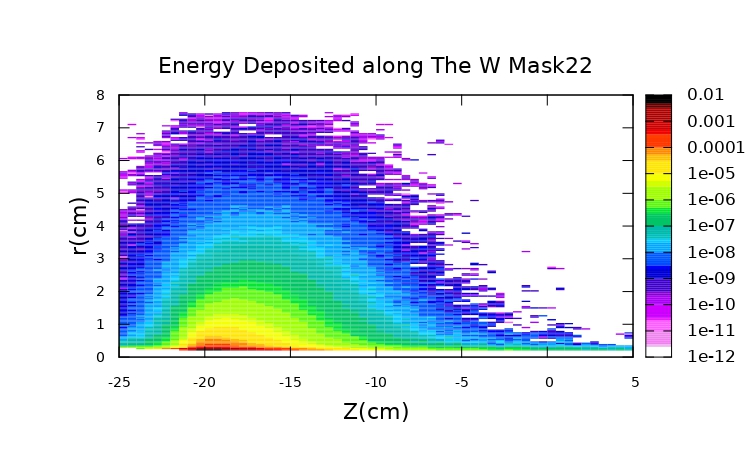}
\label{fig:graph9}
\end{subfigure}

\begin{subfigure}{0.5\textwidth}
\centering
\includegraphics[scale=0.7]{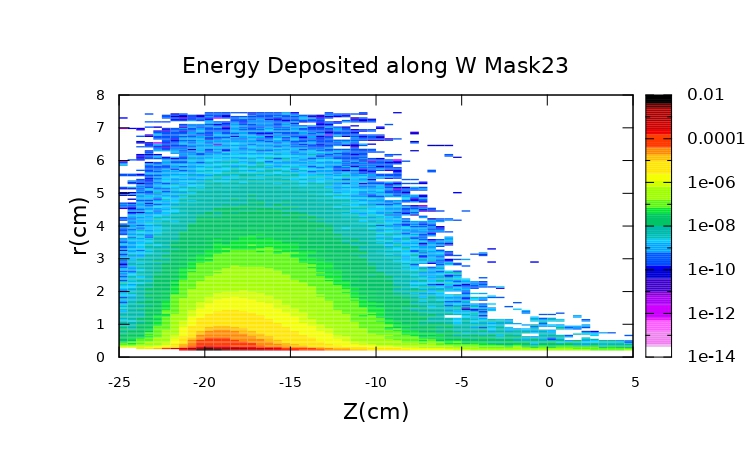}
\label{fig:graph9}
\end{subfigure}
F

\caption{Distribution of the energy deposited along the tungsten photon mask number 15, 22 and 23 from the top to the bottom, respectively, in 2D. }
\label{mul_images7}
\end{figure}

\begin{figure}[ht]
\centering

\begin{subfigure}{0.5\textwidth}
\centering
\includegraphics[scale=0.7]{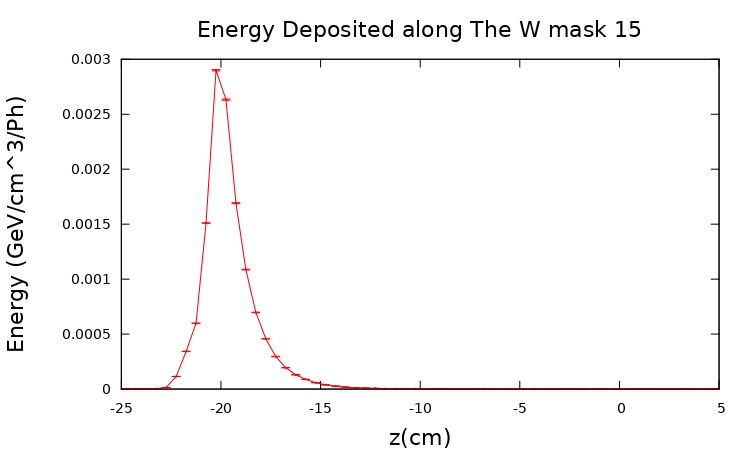}
\label{fig:graph8}
\end{subfigure}

\begin{subfigure}{0.5\textwidth}
\centering
\includegraphics[scale=0.7]{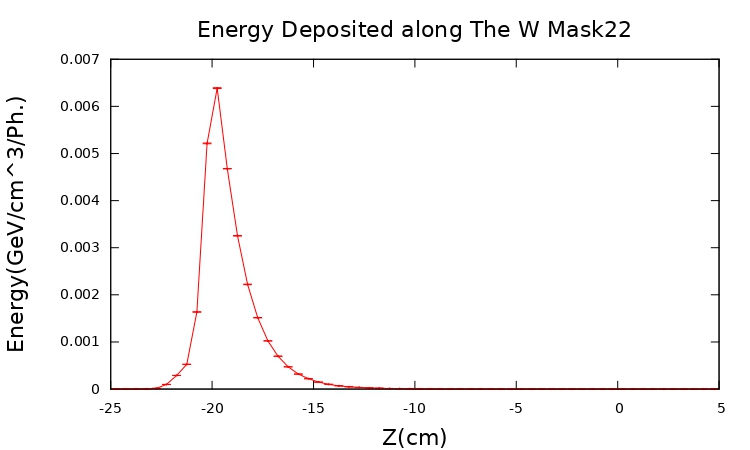}
\label{fig:graph9}
\end{subfigure}

\begin{subfigure}{0.5\textwidth}
\centering
\includegraphics[scale=0.7]{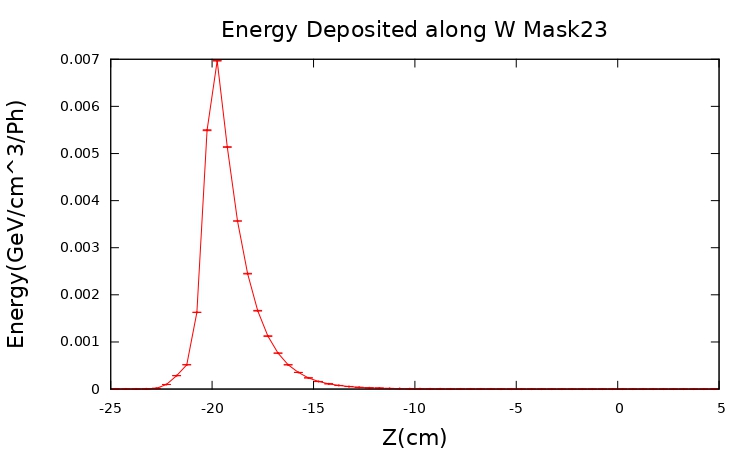}
\label{fig:graph9}
\end{subfigure}

\caption{Distribution of the energy deposited along the tungsten photon mask number 15, 22 and 23 from the top to the bottom, respectively, in 1D.}
\label{mul_images8}
\end{figure}

The last  candidate is tungsten. Tungsten has the highest density and melting point and the lowest radiation length compared to other candidates. The energy deposited at tungsten mask number 15, 22 and 23 are shown in figure \ref{mul_images7} and \ref{mul_images8} in 2D and 1D, respectively.

Table \ref{table16} shows the PEDD, maximum temperature rise and the power stopped by tungsten masks. It is clear that the PEDD at mask number 22 is lower than that at mask 23. The maximum temperature is 75.14 and 90.22 K/pulse at mask 22 and 23, respectively. The tungsten mask can stop up to 99.5 $\%$ of the deposited power.

\begin{table}[h]
\caption[ILC undulator parameters]{PEDD, maximum temperature rise and power stopped by the tungsten mask number 22 and 23.}
\centering 
\begin{tabular}{l*{6}{l}r} 
\hline\hline 

Mask ID &  Power   &    Average incident   & PEDD& Maximum  & Stopped \\
 &   Deposited  &     photon energy  &  &  temperature &  Power\\
 &  [W]   &    [MeV]   & [J/(g*pulse)]& [K/pulse]  & $\%$ \\
\hline 

22 & 270 & 1.77  &  10.07      &      75.14    &              99.5             \\
23 & 335 & 2.01  &   12.09      &        90.22   &              99.5               \\

\hline 
\end{tabular}
\label{table16} 
\end{table}

\newpage

\newpage

\newpage

\newpage


\newpage
\subsection{Electromagnetic Shower}

Since photons lose energy when they pass through a material,  secondary particles will be produced. 
The main secondary particles produced are positrons, electrons and photons. These secondary particles will again lose energy through the material via radiation processes as well as ionization. Secondaries that have enough energy to escape from the mask may deposit on the undulator vacuum. In this part power of secondaries that leaves the mask will be investigated.  \\

\begin{figure}[h]
\centering
\includegraphics[scale=1.5]{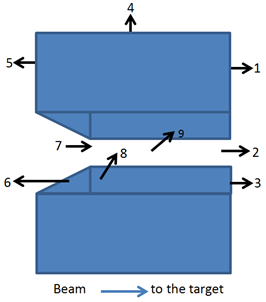}
\caption{Possible directions of secondary particles leaving a photon mask.}
\label{fig:graph90}
\end{figure}

Figure \ref {fig:graph90} shows that there are 9 possible directions that secondary particles can leave the mask to. We discuss these 9 directions, which we shall refer to as Arrow 1, Arrow 2, Arrow 3, Arrow 4, Arrow 5, Arrow 6, Arrow 7, Arrow 8 and Arrow 9: 
\begin{description}
\item[$\bullet$ Arrow 1]  represents the direction of the secondaries that leave the mask in a forward direction within the mask outer radius between 0.585 cm and 15 cm.
\item[$\bullet$ Arrow 2]  represents the direction of the secondaries leaving the hole of the mask in a forward direction .
\item[$\bullet$ Arrow 3]  represents the direction of the secondaries leaving the mask outer radius between 0.22 cm and 0.585 cm in a forward direction.
\item[$\bullet$ Arrow 4]  represents the direction of the secondaries leaving the mask in a radial direction.
\item[$\bullet$ Arrow 5]  represents the direction of the secondaries that leave the mask in a backward direction within mask outer radius between 0.585 cm and 15 cm.
\item[$\bullet$ Arrow 6]  represents the direction of the secondaries leaving the part of the tapered mask in a backward direction.
\item[$\bullet$ Arrow 7] represents the direction of secondaries leaving the tapered part and entering the hole of the mask.
\item[$\bullet$ Arrow 8] represents the direction of secondaries leaving the mask and entering the mask hole.
\item[$\bullet$ Arrow 9] represents the direction of secondaries leaving the hole mask and entering the mask medium again.

\end{description}

The power leaving mask 22  is discussed. The reason of choosing this mask is due to the fact that  mask 22 is the last mask inside the undulator. The amount of leaving power from this mask is the highest amount compared with previous masks. The power leaving from this mask may be  deposited in the next cryomodules.


\subsubsection{Spectrums of Secondary Particles Leaving the Materials}

 In this section, the spectrums of secondary particles leaving the copper, iron and tungsten mask 22 in a forward direction is discussed. 
Figure 10 shows spectrums of photons leaving the copper, iron and tungsten mask 22 in a forward direction. In this figure, it shows the spectrum of photons leaving copper, iron and tungsten mask in Arrow 3, 2 and 1 from the top to the bottom. It is clear that in all these Arrows, photons still have enough energy to escape from the mask, although there are differences in spectrums due to different materials. 

The spectrums of electrons leaving copper, iron and tungsten mask in Arrow 3, 2 and 1 from the top to the bottom is shown in figure 11. It is clear that tungsten mask can kill electrons in Arrow 1, in contrast electrons, in Arrow 2 and 3, still have enough energy to escape from the tungsten mask 22. In case of the copper and iron mask 22, electrons have enough energy to leave the mask 22 in all three arrows.

\begin{figure}[h]
  \begin{flushleft}
  \begin{tabular}{llcr}
&$~~~~~$Cupper&$~~~~~$Iron&
$~~~$Tungsten$~~~~~~$\\
 \begin{rotate}{90}
$~~~~$ Arrow 3
\end{rotate}&
\includegraphics*[width=50mm,height=35mm,angle=00]{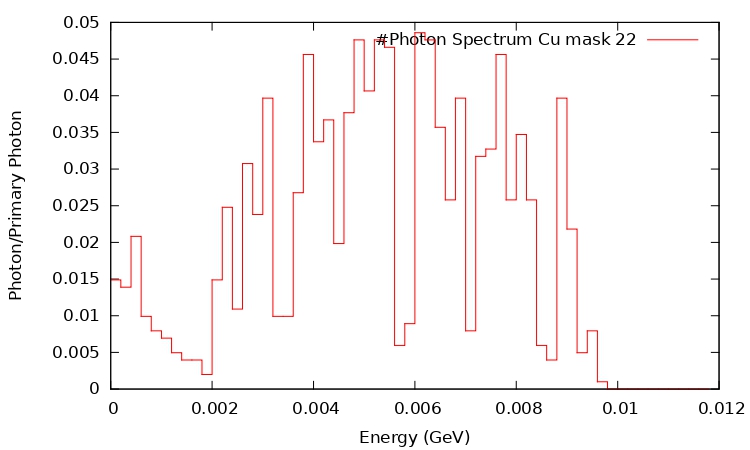}&
 \includegraphics*[width=50mm,height=35mm,angle=00]{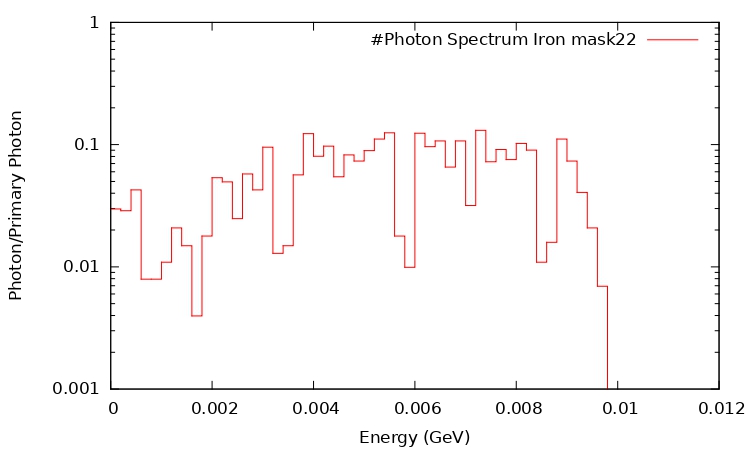}&
 \includegraphics*[width=50mm,height=35mm,angle=00]{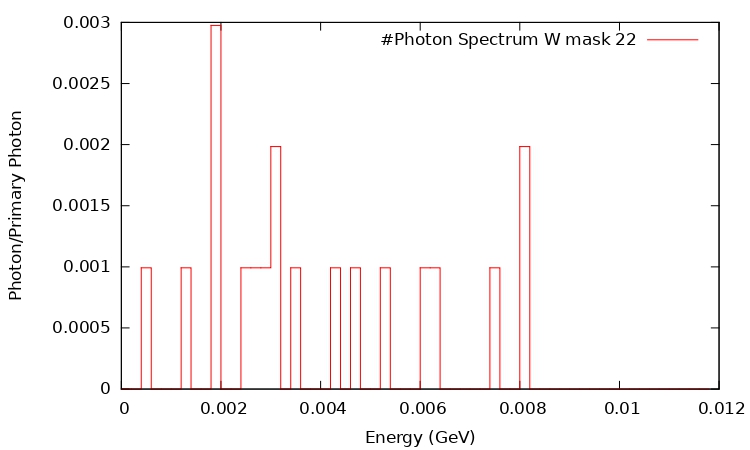}\\
 \begin{rotate}{90}
$~~~~$  Arrow 2 
\end{rotate}&
 \includegraphics*[width=50mm,height=35mm,angle=00]{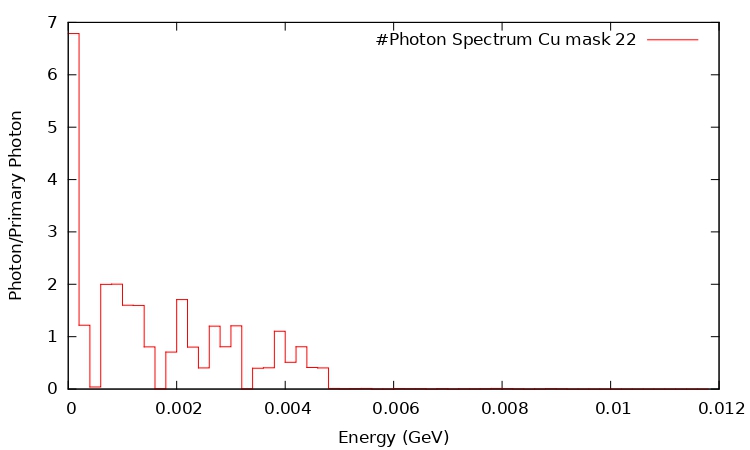}&
 \includegraphics*[width=50mm,height=35mm,angle=00]{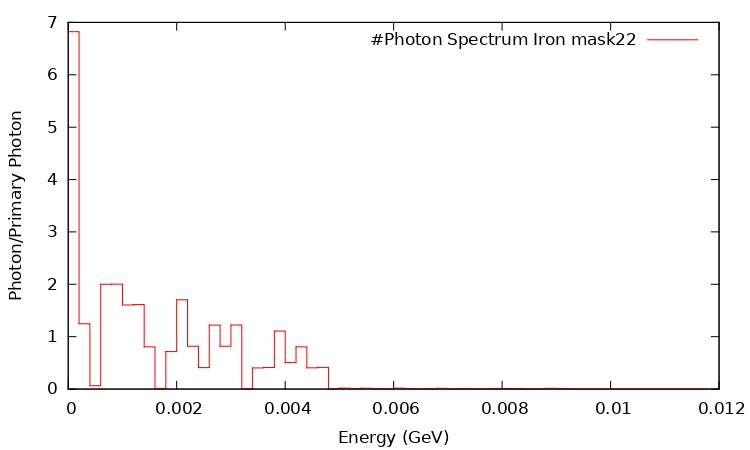}&
 \includegraphics*[width=50mm,height=35mm,angle=00]{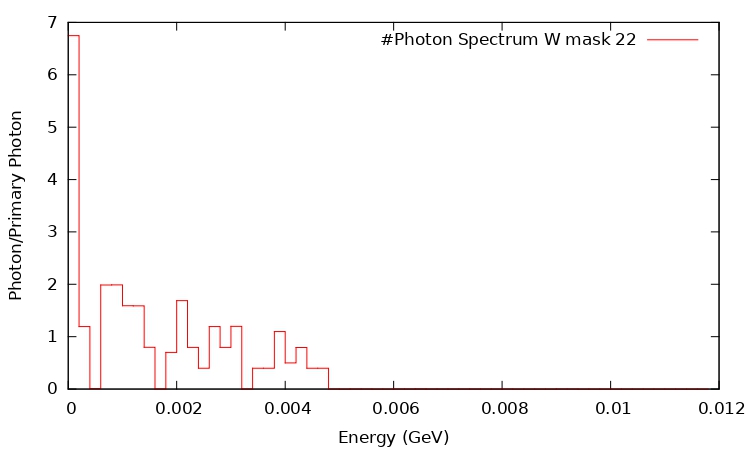}\\
 \begin{rotate}{90}
$~~~~$  Arrow 1
\end{rotate}&
 \includegraphics*[width=50mm,height=35mm,angle=00]{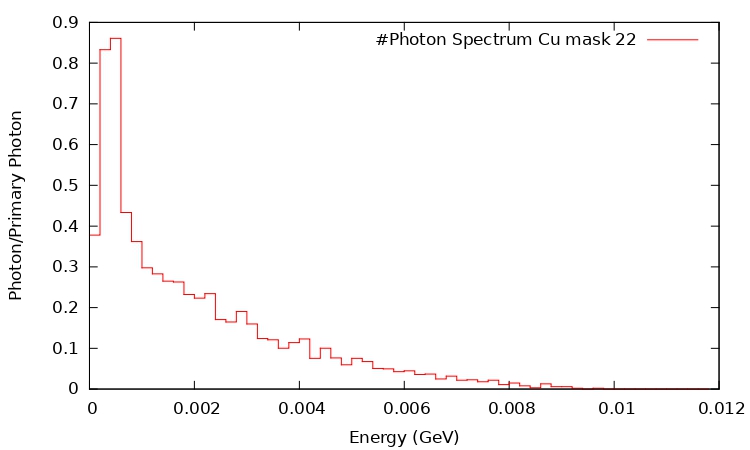}&
 \includegraphics*[width=50mm,height=35mm,angle=00]{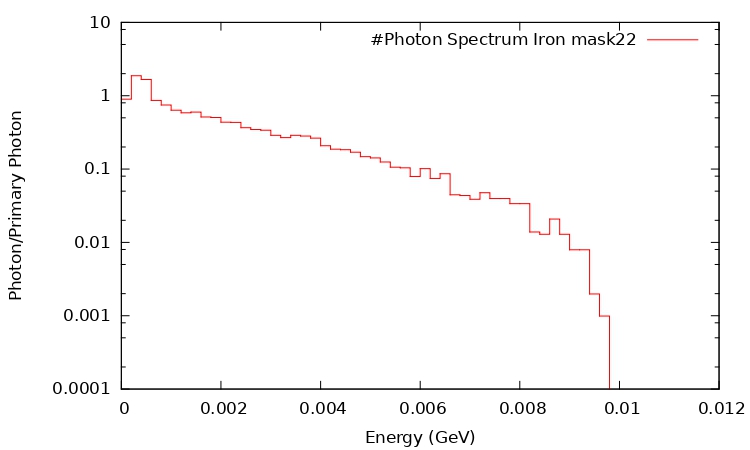}&
 \includegraphics*[width=50mm,height=35mm,angle=00]{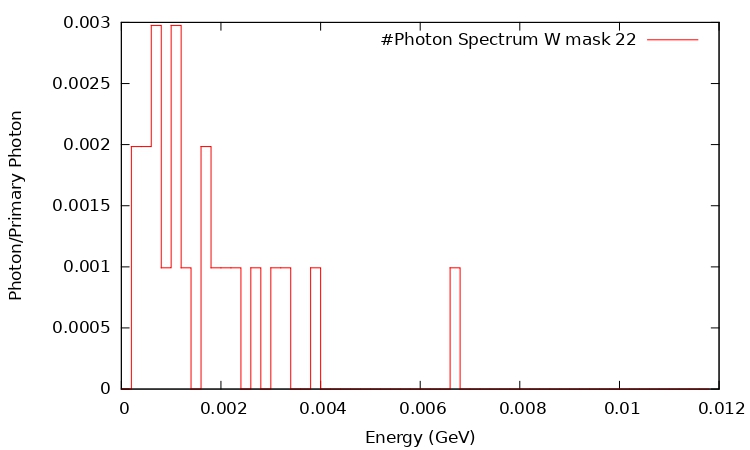}\\
\end{tabular}
  \caption{Spectrums of photons leaving tungsten, iron and copper mask 22 through  Arrow 3, 2 and 1 from the top to the bottom, respectively. }
\end{flushleft}
  \label{mul_PhotonSpecs}
\end{figure}


\begin{figure}[h]
  \begin{flushleft}
  \begin{tabular}{llcr}
&$~~~~~$Cupper&$~~~~~$Iron&
$~~~$Tungsten$~~~~~~$\\
 \begin{rotate}{90}
$~~~~$ Arrow 3
\end{rotate}&
\includegraphics*[width=50mm,height=35mm,angle=00]{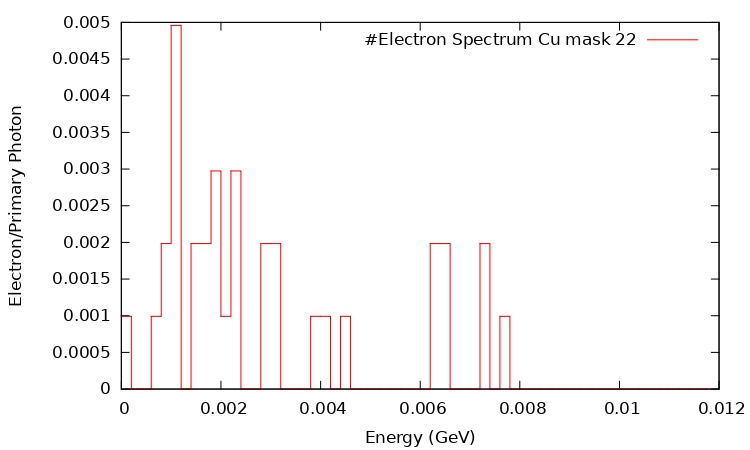}&
 \includegraphics*[width=50mm,height=35mm,angle=00]{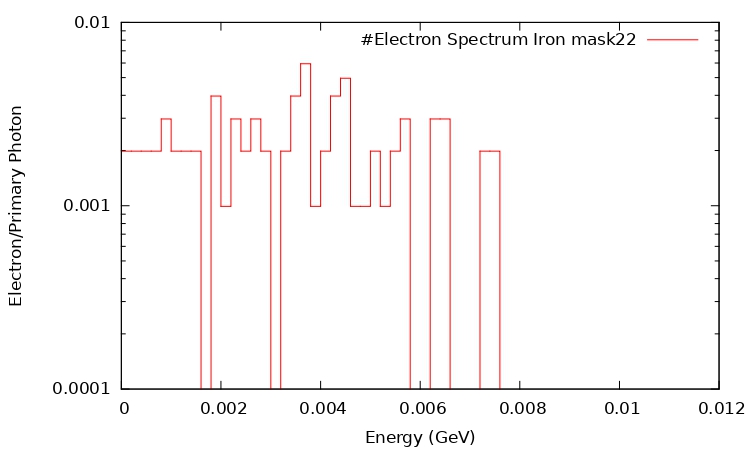}&
 \includegraphics*[width=50mm,height=35mm,angle=00]{PhotSpeW22A3.png}\\
 \begin{rotate}{90}
$~~~~$  Arrow 2 
\end{rotate}&
 \includegraphics*[width=50mm,height=35mm,angle=00]{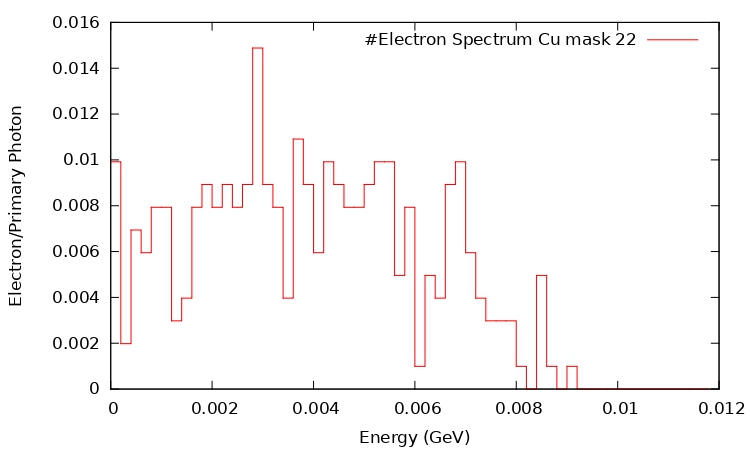}&
 \includegraphics*[width=50mm,height=35mm,angle=00]{PhotSpeFe22A2.png}&
 \includegraphics*[width=50mm,height=35mm,angle=00]{PhotSpeW22A2.png}\\
 \begin{rotate}{90}
$~~~~$ Arrow 1
\end{rotate}&
 \includegraphics*[width=50mm,height=35mm,angle=00]{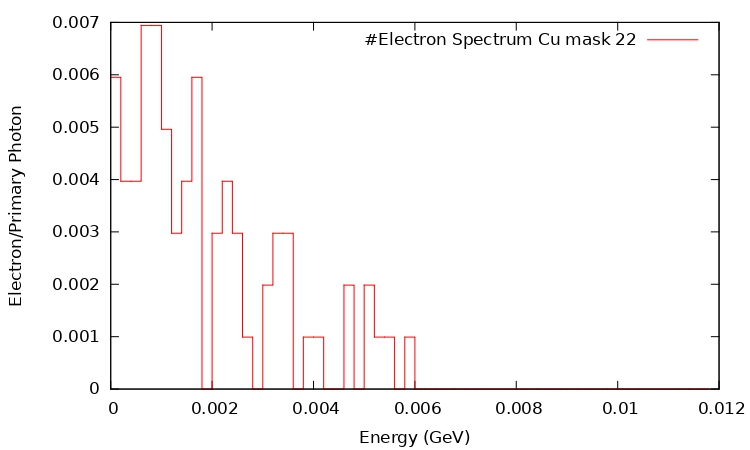}&
 \includegraphics*[width=50mm,height=35mm,angle=00]{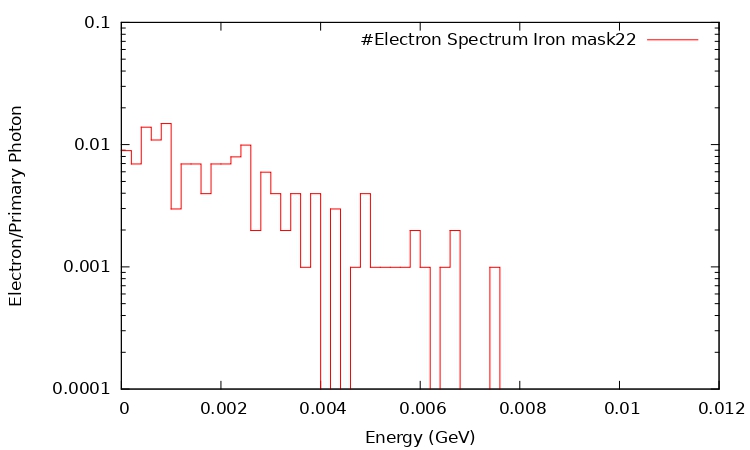}&
\end{tabular}
  \caption{Spectrums of electrons leaving tungsten, iron and copper mask 22 through  Arrow 3, 2 and 1 from the top to the bottom, respectively.}
\end{flushleft}
  \label{mul_PhotonSpecs}
\end{figure}


\subsubsection{Power Leaving The Copper Mask}

In the previous section, the spectrums of secondary particles (including photons and electrons) leaving the copper, iron and tungsten mask 22 in a forward direction was discussed. In this section, the power of secondary particles leaving mask 22 with three different materials in all directions is investigated.

\begin{description}
\item[$\bullet$ Power Leaving The Copper Mask 22]
\end{description}

Table \ref{tableCu22} shows the average energy and the power of each secondary particle at each arrow from copper mask 22. The secondary particles including photon, electron and positron leave the mask to all directions, except there 
are no positrons leaving the Cu mask 22 to Arrow 5. The total power that leaves the Cu mask 22 is 4.55 W. The power that leaves the mask to forward direction (total power of Arrow 1, Arrow 2 and Arrow3) is 2.18 W. In section 5.5, the power deposition of secondary particles on the undulator vacuum will be discussed.

\begin{table}[htbp]
  \centering
  \caption{Power of secondary particles leaving the copper mask 22 to all directions.}
    \begin{tabular}{lrrrr}
\hline\hline 
          & \multicolumn{1}{l}{Average Energy (MeV)} & \multicolumn{1}{l}{Particle/Primary Photon} & \multicolumn{1}{l}{Total Power (W)} & \multicolumn{1}{l}{Total Power (W)} \\
 &&&& ($\gamma$, e-,e+) \\
\hline 
\textbf{Arrow 1} &       &       &       &  \\
\hline 
    Photon & 2     & 1.37E-03 & 4.57E-01 & 0.46 \\
    Electron & 1.85  & 1.47E-05 & 4.53E-03 &  \\
    Positron & 2.74  & 9.92E-07 & 4.53E-04 &  \\
\hline 
    \textbf{Arrow 2} &       &       &       &  \\
\hline 
    Photon & 1.63  & 5.42E-03 & 1.47E+00 & 1.52 \\
    Electron & 3.92  & 5.83E-05 & 3.81E-02 &  \\
    Positron & 3.76  & 1.45E-05 & 9.08E-03 &  \\
\hline 
  \textbf{Arrow 3} &       &       &       &  \\
\hline 
    Photon & 5.29  & 2.28E-04 & 2.01E-01 & 0.20 \\
    Electron & 2.99  & 6.74E-06 & 3.36E-03 &  \\
    Positron & 3.1   & 1.39E-06 & 7.18E-04 &  \\
\hline 
    \textbf{Arrow 4} &       &       &       &  \\
\hline 
    Photon & 0.588 & 2.27E-02 & 2.23E+00 & 2.23 \\
    Electron & 0.496 & 5.63E-05 & 4.66E-03 &  \\
    Positron & 1.03  & 3.97E-07 & 6.82E-05 &  \\
\hline 
    \textbf{Arrow 5} &       &       &       &  \\
\hline 
    Photon & 0.366 & 1.97E-03 & 1.21E-01 & 0.12 \\
    Electron & 0.248 & 3.57E-06 & 1.48E-04 &  \\
    Positron & 0     & 0.00E+00 & 0.00E+00 &  \\
\hline 
    \textbf{Arrow 6} &       &       &       &  \\
\hline 
    Photon & 0.221 & 3.08E-04 & 1.14E-02 & 0.02 \\
    Electron & 0.38  & 5.85E-05 & 3.71E-03 &  \\
    Positron & 0.855 & 5.75E-06 & 8.20E-04 &  \\

\hline 
    \textbf{Arrow 7} &       &       &       &  \\
\hline 
    Photon & 1.12  & 2.73E-02 &  5.106 & 11.90 \\
    Electron & 1.93  & 1.74E-02 & 5.617 &  \\
    Positron & 2.53  & 2.80E-03 &  1.1818 &  \\
\hline 
    \textbf{Arrow 8} &       &       &       &  \\
\hline 
    Photon & 0.774 & 3.18E-01 & 41.031 & 154.038 \\
    Electron & 1.57  & 3.48E-01 & 91.24 &  \\
    Positron & 1.99  & 6.56E-02 & 21.76 &  \\
\hline 

      \textbf{Arrow 9} &       &       &       &  \\
\hline 
    Photon & 0.801 & 3.31E-01 & 44.27 & 163.413 \\
    Electron & 1.6   & 3.60E-01 & 96.21 &  \\
    Positron & 2.03  & 6.77E-02 & 22.93 &  \\

   \hline 
    \end{tabular}%
  \label{tableCu22}%
\end{table}%





\begin{description}
\item[$\bullet$ Power Leaving The Iron Mask 22]
\end{description}

Table \ref{tableFe22} shows the average energy and the power of each single secondary particle at each arrow from Fe Mask 22. The secondary particles including photons, electrons and positrons are travelling in all directions around the mask, except there are no positrons travelling to Arrow 5. The total power that leaves the Fe mask 22 in all possible directions is 6.64 W. Only 3.07 W will go in a forward direction. This amount of power (3.07 W) needs to be discussed to find out how much power will be deposited in next cryomodule and how much power will go through the undulator vacuum. In section 5.5, the power deposition of secondary particles on the undulator vacuum will be discussed.

\begin{table}[htbp]
  \centering
  \caption{Power of secondary particles leaving the iron mask 22 to all directions.}
    \begin{tabular}{lrrrr}
\hline\hline
          & \multicolumn{1}{l}{Average Energy (MeV)} & \multicolumn{1}{l}{Particle/Primary Photon} & \multicolumn{1}{l}{Total Power (W)} & \multicolumn{1}{l}{Total Power (W)  } \\
  &&&& ($\gamma$, e-,e+) \\
\hline   
 \textbf{Arrow 1} &       &       &       &  \\
\hline     
  Photon & 1.98  & 2.86E-03 & 9.45E-01 & 9.57E-01 \\
    Electron & 2.06  & 3.00E-05 & 1.03E-02 &  \\
    Positron & 1.98  & 3.77E-06 & 1.24E-03 &  \\
\hline   
    \textbf{Arrow 2} &       &       &       &  \\
\hline   
    Photon & 1.64  & 5.47E-03 & 1.50E+00 & 1.58E+00 \\
    Electron & 3.87  & 1.08E-04 & 6.95E-02 &  \\
    Positron & 3.65  & 3.03E-05 & 1.85E-02 &  \\
\hline   
\textbf{Arrow 3} &       &       &       &  \\
\hline   
    Photon & 5.58  & 5.66E-04 & 5.27E-01 & 5.37E-01 \\
    Electron & 3.49  & 1.51E-05 & 8.78E-03 &  \\
    Positron & 2.59  & 3.37E-06 & 1.46E-03 &  \\
\hline   
 \textbf{Arrow 4} &       &       &       &  \\
\hline   
    Photon & 0.543 & 3.72E-02 & 3.37E+00 & 3.38E+00 \\
    Electron & 0.494 & 8.85E-05 & 7.29E-03 &  \\
    Positron & 1.14  & 9.92E-07 & 1.89E-04 &  \\
\hline   
\textbf{Arrow 5} &       &       &       &  \\
\hline   
    Photon & 0.315 & 3.08E-03 & 1.62E-01 & 1.62E-01 \\
    Electron & 0.361 & 3.57E-06 & 2.15E-04 &  \\
    Positron & 0     & 0.00E+00 & 0.00E+00 &  \\
\hline   
\textbf{Arrow 6} &       &       &       &  \\
\hline   
    Photon & 0.212 & 3.30E-04 & 2.38E-02 & 2.82E-02 \\
    Electron & 0.432 & 5.36E-05 & 3.86E-03 &  \\
    Positron & 0.904 & 3.57E-06 & 5.38E-04 &  \\

\hline   
 \textbf{Arrow 7} &       &       &       &  \\
\hline   
    Photon & 1.1   & 2.61E-02 & 4.78 & 10.91 \\
    Electron & 1.91  & 1.62E-02 & 5.17 &  \\
    Positron & 2.53  & 2.28E-03 &  0.96 &  \\
\hline   
 \textbf{Arrow 8} &       &       &       &  \\
\hline   
    Photon & 0.775 & 3.22E-01 & 41.68 & 155.40 \\
    Electron & 1.58  & 3.54E-01 & 93.22 &  \\
    Positron & 2.01  & 6.12E-02 &  20.50 &  \\
\hline   
 \textbf{Arrow 9} &       &       &       &  \\
\hline   
    Photon & 0.779 & 3.35E-01 & 43.54 & 163.07 \\
    Electron & 1.61  & 3.65E-01 &  98.11 &  \\
    Positron & 2.04  & 6.29E-02 & 21.42 &  \\

\hline   
    \end{tabular}%
  \label{tableFe22}%
\end{table}%



\begin{description}
\item[$\bullet$ Power Leaving The Tungsten Mask 22]
\end{description}

Table \ref{tableW22} shows the average energy and the power of each secondary particle at each arrow from W Mask 22. It is clear that photons are travelling to all directions around the mask. While tungsten mask 22 can kill electrons and positrons such as which escape from the W mask to Arrow 1. The total power that leaves the W mask 22 to all possible directions is 1.52 W. Most of this power (1.43 W) will travel within Arrow 2. This amount of power (1.43 W) will be discussed to find out how much power will be deposited in next cryomodules and how much power will go through the undulator vacuum.  In section 5.5, the power deposition of secondary particle on the undulator vacuum will be discussed.

\begin{table}[htbp]
  \centering
  \caption{Power of secondary particles leaving the tungsten mask 22 to all directions.}
    \begin{tabular}{lrrrr}
\hline\hline   
          & \multicolumn{1}{l}{Average Energy (MeV)} & \multicolumn{1}{l}{Particle/Primary Photon} & \multicolumn{1}{l}{Total Power (W)} & \multicolumn{1}{l}{Total Power (W)} \\
 &&&&($\gamma$, e-,e+) \\
\hline   
    \textbf{Arrow 1} &       &       &       &  \\
\hline   
    Photon & 1.75  & 4.36E-06 & 1.27E-03 & 1.274E-03 \\
    Electron & 0     & 0     & 0     &  \\
    Positron & 0     & 0     & 0     &  \\
\hline   
    \textbf{Arrow 2} &       &       &       &  \\
\hline   
    Photon & 1.6   & 5.34E-03 & 1.43E+00 & 1.428E+00 \\
    Electron & 2.92  & 4.76E-06 & 2.32E-03 &  \\
    Positron & 3.06  & 1.98E-06 & 1.01E-03 &  \\
\hline   
    \textbf{Arrow 3} &       &       &       &  \\
\hline   
    Photon & 3.98  & 3.77E-06 & 2.50E-03 & 2.532E-03 \\
    Electron & 0.9   & 1.98E-07 & 2.98E-05 &  \\
    Positron & 0     & 0     & 0     &  \\
\hline   
    \textbf{Arrow 4} &       &       &       &  \\
\hline   
    Photon & 1.41  & 1.30E-04 & 3.06E-02 & 3.092E-02 \\
    Electron & 1.17  & 1.39E-06 & 2.71E-04 &  \\
    Positron & 0     & 0     & 0     &  \\
\hline   
    \textbf{Arrow 5} &       &       &       &  \\
\hline   
    Photon & 1.02  & 1.11E-04 & 1.90E-02 & 1.898E-02 \\
    Electron & 0.283 & 1.98E-07 & 9.36E-06 &  \\
    Positron & 0     & 0     & 0     &  \\
\hline   
    \textbf{Arrow 6} &       &       &       &  \\
\hline   
    Photon & 0.419 & 2.30E-04 & 1.61E-02 & 3.220E-02 \\
    Electron & 0.415 & 1.77E-04 & 1.22E-02 &  \\
    Positron & 0.815 & 2.86E-05 & 3.88E-03 &  \\

\hline   
    \textbf{Arrow 7} &       &       &       &  \\
\hline   
    Photon & 1.19  & 3.8679533E-02 &  7.677577864 & 21.2939983      \\
    Electron & 1.88  & 3.0997144E-02 & 9.720208404 &       \\
    Positron & 2.57  & 9.0889437E-03 & 3.89621203 &      \\

\hline   
    \textbf{Arrow 8} &       &       &       &  \\
\hline   
    Photon & 0.729 & 2.6074940E-01 &  31.70639694 & 134.1688465   \\
    Electron & 1.44  & 2.9984570E-01 & 72.02053837 &        \\
    Positron & 1.87  & 9.7596504E-02 &  30.44191114 &        \\

\hline   
    \textbf{Arrow 9} &       &       &       &  \\
\hline   
    Photon & 0.791 & 2.8073750E-01 & 37.04016887 & 150.4617071   \\
    Electron & 1.51  & 3.1672230E-01 & 79.77221226 &         \\
    Positron & 1.96  & 1.0292580E-01 & 33.64932594 &         \\

\hline   
    \end{tabular}%
  \label{tableW22}%
\end{table}%






\subsection{Power Deposited at the Undulator vacuum}

The power deposition due to the primary beam along the helical undulator vacuum at ILC-250GeV was studied on \cite {alharbi2019energy}. In that study it was demonstrated that 23 photon masks are needed to be inserted along the undulator line to keep the power deposition below the acceptable limit. In addition it was assumed that the photon masks are ideal but in reality it is impossible to have ideal photon masks. Therefore in this section the energy deposition due to the secondary particles on undulator vacuum are studied for the ideal case.

\begin{description}
\item[$\bullet$ Power Deposition of the Primary Beam at the Undulator vacuum]
\end{description}

\begin{figure}[h]
\centering
\includegraphics[scale=0.7]{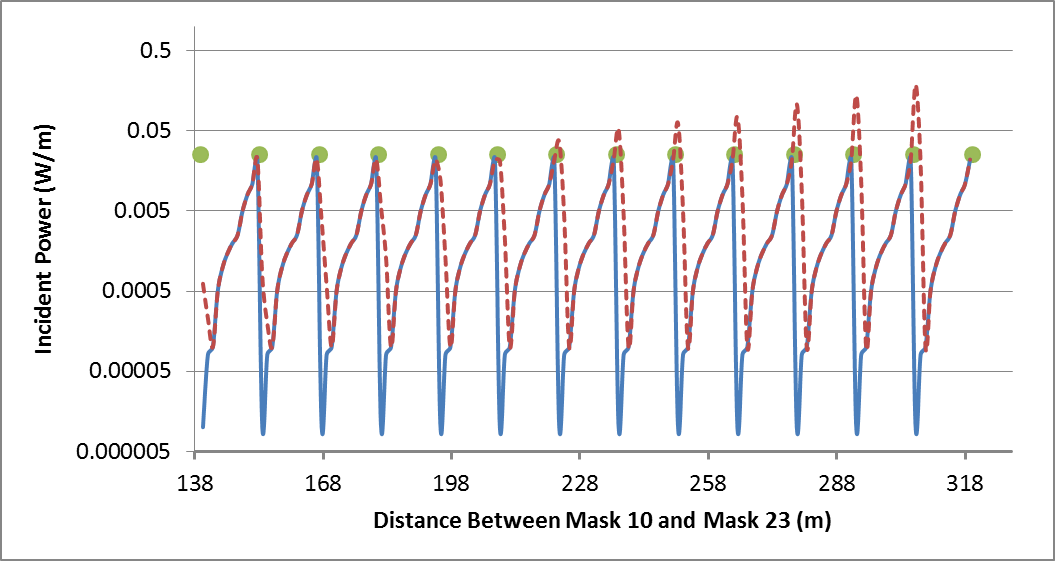}
\caption{Power deposition on the undulator vacuum between Mask 10 and Mask 23. The blue line represents the power deposition of primary photon on the undulator vacuum and the dashed red line represents the power deposition of secondary particles. The green points represent the position of the masks}
\label{fig:graph19}
\end{figure}

The blue line in figure \ref{fig:graph19} shows the power deposition of the primary photon in the undulator vacuum between Mask 10 and Mask 23. Since it is an ideal helical undulator, the peak and minimum of the primary beam power between Mask 10 and Mask 23  are almost the same. The peak of the primary beam power is 0.022 W/m and it is at the last module before the next mask. The minimum power is $1\times10^{-5}$ W/m and it is at the beginning of the first module after each mask.

\begin{description}
\item[$\bullet$ Power Deposition of Secondary Particles at the Undulator vacuum]
\end{description}

The power deposition of secondary particles at undulator vacuum was simulated using FLUKA. In section 5.4 three different materials including copper, iron and tungsten were invistagated. In this section the mask material used is copper. The peaks of the power deposited of secondaries depend on the incident power of the primary beam. So that the power leaving mask is different and that  leads to differences on the peaks of the power deposition of secondary particles in the undulator vacuum. The dashed red line on figure \ref{fig:graph19} represents the power deposition of secondary particles in the undulator vacuum between Mask 10 and Mask 23. 

Since the Mask 23 is at the exit of the undulator, thus the power leaving Mask 23 is not important to our study. As it can be seen on figure \ref{fig:graph19} the highest peak of power deposition of secondary particles is immediately after Mask 22.\\

Table \ref{tab:addlabel} summrizes the power deposition of the primary beam and the total deposition power between Mask 21 and Mask 23. The total deposition power means the power deposition of the primary beam plus the power deposition of secondary particles.

\begin{table}[htbp]
  \centering
  \caption{Power deposition of the primary beam and the total power deposition between Mask 21 and Mask 23.}
    \begin{tabular}{llrr}
\hline   \hline   
          &       & \multicolumn{1}{l}{Power Deposition of } & \multicolumn{1}{l}{Total Power Depsotion at } \\
          &       & \multicolumn{1}{l}{ the Primary Beam (W/m)} & \multicolumn{1}{l}{Undulator vacuum (W/m)} \\
\hline   \hline   

    Mask 21 & Distance (m) &       &  \\
\hline  
          & \multicolumn{1}{r}{1} & 0.00001 & 0.14018 \\
          & \multicolumn{1}{r}{2} & 0.00008 & 0.01148 \\
          & \multicolumn{1}{r}{3} & 0.0001 & 0.0001 \\
          & \multicolumn{1}{r}{4} & 0.0005 & 0.0005 \\
          & \multicolumn{1}{r}{5} & 0.001 & 0.001 \\
          & \multicolumn{1}{r}{6} & 0.0015 & 0.0015 \\
          & \multicolumn{1}{r}{7} & 0.002 & 0.002 \\
          & \multicolumn{1}{r}{8} & 0.0025 & 0.0025 \\
          & \multicolumn{1}{r}{9} & 0.005 & 0.005 \\
          & \multicolumn{1}{r}{10} & 0.008 & 0.008 \\
          & \multicolumn{1}{r}{11} & 0.011 & 0.011 \\
          & \multicolumn{1}{r}{12} & 0.022 & 0.022 \\
\hline  
    Mask 22 & Distance (m) &       &  \\
\hline   
          & \multicolumn{1}{r}{1} & 0.00001 & 0.19001 \\
          & \multicolumn{1}{r}{2} & 0.00008 & 0.02008 \\
          & \multicolumn{1}{r}{3} & 0.0001 & 0.0001 \\
          & \multicolumn{1}{r}{4} & 0.0005 & 0.0005 \\
          & \multicolumn{1}{r}{5} & 0.001 & 0.001 \\
          & \multicolumn{1}{r}{6} & 0.0015 & 0.0015 \\
          & \multicolumn{1}{r}{7} & 0.002 & 0.002 \\
          & \multicolumn{1}{r}{8} & 0.0025 & 0.0025 \\
          & \multicolumn{1}{r}{9} & 0.005 & 0.005 \\
          & \multicolumn{1}{r}{10} & 0.008 & 0.008 \\
          & \multicolumn{1}{r}{11} & 0.011 & 0.011 \\
          & \multicolumn{1}{r}{12} & 0.022 & 0.022 \\
\hline   
    Mask 23 & Exit Undulator &       &  \\
\hline  
    \end{tabular}%
  \label{tab:addlabel}%
\end{table}%

\section{Conclusion}

We have modelled a possible photon mask geometry with high absorption efficiency for three different materials including copper, iron and tungsten at the ILC positron source. Copper and iron are stopped 97.5\% and 98.5\% of the incident power, respectively. In contrast, tungsten can stop up to 99.5\% of the incident power. 

PEDD and maximum temperature increase studies showed that for ideal helical undulator, the masks are safe. For example, the PEDD at copper mask 23 is 8.18 J/(g*pulse), and maximum temperature rise is 21.25 K/pulse. According to the radiation length, the photon mask can be shorter by using tungsten.

It has been shown that secondary particles will deposit in the undulator vacuum. For the ideal undulator, the energy deposition in the undulator vacuum due to the synchrotron radiation and secondary particles is below the acceptable limit. For instance, the peak  of the energy deposition immediately after Mask 22 is 0.19 W/m.   

The next step will be to simulate the photon spectrum for a non-ideal undulator and to finalise the design of the photon mask.

\newpage

\begin{description}
\fontsize{13}{13}\selectfont
\item[$\bullet$ Acknowledgments]
\end{description}
\fontsize{12}{13}\selectfont
GMP was partially supported by the Deutsche Forschungsgemeinschaft (DFG, German Research Foundation) under Germany's Excellence Strategy -  EXC 2121 „Quantum Universe“ - 390833306.


\bibliographystyle{vancouver}
\bibliography{LCWS2019}

\end{document}